\documentclass[12pt]{article}

\usepackage{amsmath}
\usepackage{mathtools}
\usepackage{dashbox}
\usepackage{a4wide}
\usepackage{epsfig}
\usepackage{theorem}
\usepackage{amssymb}
\usepackage{bbm}
\usepackage[T1]{fontenc}
\usepackage[latin1]{inputenc}
\usepackage{cancel}
\usepackage{blkarray}
\usepackage{multirow}

\newcommand{\cD}{{\cal D}}

\newcommand{\cH}{{\cal H}}

\newcommand{\bi}{\bigskip}
\newcommand{\no}{\noindent}
\newcommand{\bea}{\begin{eqnarray}}
\newcommand\dboxed[1]{\dbox{\ensuremath{#1}}}
\newcommand{\eea}{\end{eqnarray}}

\newcommand{\be}{\begin{equation}}
\newcommand{\ee}{\end{equation}}

\newcommand{\lk}{\left(}\usepackage{multirow}

\newcommand{\sli}{\sum\limits}

\newcommand{\vx}{\vec{x}}
\newcommand{\vz}{\vec{z}}
\newcommand{\vC}{\vec{C}}
\newcommand{\vp}{\vec{p}}

\newcommand{\vA}{\vec{A}}
\newcommand{\vB}{\vec{B}}

\newcommand{\vq}{\vec{q}}

\newcommand{\vL}{\vec{L}}
\newcommand{\vr}{\vec{r}}
\newcommand{\il}{\int\limits}
\newcommand{\vv}{\vec{v}}

\newcommand{\vK}{\vec{K}}

\usepackage{amsmath}
\usepackage{a4wide}
\usepackage{epsfig}
\usepackage{theorem}
\usepackage{amssymb}
\usepackage{bbm}
\usepackage[T1]{fontenc}
\usepackage[latin1]{inputenc}

\begin{document}

\title{Classical and Quantum Mechanics with Lie Brackets and Pseudocanonical Transformations}

\author{W.~Dittrich\\
Institut f\"ur Theoretische Physik\\
Universit\"at T\"ubingen\\
Auf der Morgenstelle 14\\
D-72076 T\"ubingen\\
Germany\\
electronic address: qed.dittrich@uni-tuebingen.de
}
\date{\today}

\maketitle
\bi

\no

\begin{abstract}
We emphasize the usefulness of the Lie brackets in the context of classical and quantum mechanics. 
By way of examples we show that many dynamical systems, especially the ones with (gauge) constraints, can equally be treated in their 
time development with non-canonical variables and Hamiltonians.  After a short presentation of the Lie bracket algebra and 
treating some easier standard problems with the Lie bracket techniques, we concentrate mainly on charged particles with gauge 
constraint in a constant external magnetic field.
Since most of our quantum field theories are meanwhile considered effective, we have purposely treated our final problems with 
$c$-number instead of field -operator Lagrangians. The van Vleck determinant, which is exact for our problems, is employed to 
calculate the $c$-number Feynman-Schwinger propagation function. There is no need for operators or renormalization. 
In particular, the non-relativistic propagator in $2+1$ dimensions and the more complicated one in $3+1$
dimensions are presented in all their glorious detail. On the more editorial side: we have dispensed with numerating the various 
problems. They are not so much disjoint that they needed an extra title. Also, the article is written in a self-consistent
way, meaning one should be able to read it without time-consuming research in textbooks and journals - 
with a few exceptions, in particular Schwinger's paper \cite{2}, 
which is the most-cited paper in modern quantum-field-theory physics. Most of the prerequisites for reading the present 
article can be found in extenso in \cite{1}.

\end{abstract}

%\section{}

We know that in Hamiltonian systems a dynamic function $f(q,p)$ develops in time according to
\be
\label{1}
\dot{f} = [f,H]_{P.B} \, .							
\ee
In classical mechanics, we are used to studying the time development of a physical system by employing the Hamiltonian and the 
Poisson brackets $(P.B.)$. 
But one can extend the latter, using the so-called Lie brackets $(L.B.)$ to great advantage.
A Lie algebra is an algebraic structure in which the connections between its elements  
are determined with the help of $L.B$. 
In order to do this, we begin with the set $\cD$
of all dynamic functions $A(q,p), B(q,p), \ldots \epsilon \cD$ and require that $\alpha A, \alpha A + \beta B, AB$ and $A^{-1}
(\alpha, \beta$ constants) be defined and likewise should belong to $\cD$. 
When we consider elements like $A, B, \ldots$ as building blocks, then one can construct a large class of dynamic functions, 
i.e., polynomials, analytic functions, meromorphic functions, Fourier series, etc. One only gets a really new characteristic of 
$\cD$ if one imposes an algebraic structure by way of the Lie bracket $[A,B] \epsilon \cD$, where $L.B.$
satisfies the following relations:
\begin{align}
 \label{2}
{[A,B]} & = -[B,A] \, , \\
 \label{3}
 [\alpha A,B] & = \alpha [A,B] 	\, , \\
 \label{4}
 [\alpha A + \beta B,C] & =  \alpha [A,C] + \beta [B,C] \, , \\
\label{5}
					[AB,C] & = A[B,C] + B[A,C] \, .  		
 \end{align}
			The operation $L.B.$ is not associative. Instead the Jacobi identity applies:
\be
\label{6}
					[[A,B],C] = [[B,C],A] + [[C,A],B]] = 0 \, .
\ee
A well-known example is given by a 3-dimensional vector space with cross product:
\begin{align}        
				 \vA \times \vB & = - \vB \times \vA \, , \nonumber\\
				(\alpha \vA + \beta \vB) \times \vC & = \alpha \vA \times  \vC +  \beta \vB \times \vC \, , \nonumber\\
				\vA \times  (\vB \times \vC) + \vB \times (\vC \times \vA) + \vC \times  (\vA \times \vB) & = 0 \, .\nonumber 
\end{align}
The three rules (\ref{3}, \ref{4}, \ref{5}) can be realized with the help of first-order differential operators:
\be
\label{7}
[A (q, p), B] = \sli^N_{a = 1} \lk \frac{\partial A}{\partial q_a} \left[ q_a, B \right] + \frac{\partial A}{\partial p_a} [p_a, B] \right) \, .
\ee
The proof for (\ref{3}) and (\ref{4}) is trivial, as is the proof for (\ref{5}):
\begin{align}
 [AB, C] &= \frac{\partial (AB)}{\partial q} [q, C] + \frac{\partial (AB)}{\partial p} [p, C] \nonumber\\
 &= \lk \frac{\partial A}{\partial q} B + A \frac{\partial B}{\partial q} \right) [q, C] + 
 \lk \frac{\partial A}{\partial p} B + A \frac{\partial B}{\partial p} \right) [p, C] \nonumber\\
 &= A \lk \frac{\partial B}{\partial q} [q, C] + \frac{\partial B}{\partial p} [p, C] \right) + 
 B \lk \frac{\partial A}{\partial q} [q, C] + \frac{\partial A}{\partial p} [p, C] \right) \nonumber\\
 &= A [B, C] + B [A, C]  \, .\nonumber
 %\label{170-2}
\end{align}
If one iterates relation (\ref{7}), one finds:
 \begin{align}
  [A (q, p), B (q, p)] &= \sli^N_{a = 1} 
 \sli^N_{b = 1} \Bigg( \frac{\partial A}{\partial q_a} \frac{\partial B}{\partial q_b} 
 [q_a, q_b] + \frac{\partial A}{\partial q_a} \frac{\partial B}{\partial p_b} [q_a,  p_b] 
  \nonumber \\
  &  + \frac{\partial A}{\partial p_a} \frac{\partial B}{\partial q_b} [p_a, q_b] + \frac{\partial A}{\partial p_a} 
  \frac{\partial B}{\partial p_b} [p_a, p_b] \Bigg) \, . 
\label{8}
 \end{align}
The $q$'s and $p$'s here are arbitrary pairs of phase-space variables - not necessarily canonical conjugate variables as they appear 
in the Hamiltonian equations.

The quantities $[q_a, q_b], [q_a, p_b], [p_a, p_b]$
are called fundamental $L.B$. If one knows these for all $a, b = 1, \ldots N$, then one can calculate the Lie brackets in (\ref{8}).
If, however, we define the fundamental L.B. according to
\be
\label{9}
[q_i, q_j] = 0 \, , \quad [p_i, p_j] = 0 \, , \quad [q_i, p_j] = \delta_{ij} \, \quad \mbox{for all} \, i, j \, , 
\ee
the $L.B.$ go over to the $P.B.$:
\be
\label{10}
[A, B]_{q, p} = \sli^N_{i, j = 1} \lk \frac{\partial A}{\partial q_i} \frac{\partial B}{\partial p_j} - \frac{\partial A}{\partial p_i}
\frac{\partial B}{\partial q_j} \right) \, .
\ee
The variables $q_i, p_i$ are then said to be canonically conjugate.
For a Hamiltonian system $H$, the time evolutionary equations then immediately follow  from (\ref{8}) for the variables $q$ and $p$:
												\begin{align}
												 \label{11}
A = q, B = H &: \dot{q}_i = [q_i, H] = [q_i, p_j] \frac{\partial H}{\partial p_j} = \delta_{ij} \frac{\partial H}{\partial p_j}
= \frac{\partial H}{\partial p_i} \, , \\
\label{12}
A = p, B = H &: \dot{p}_i = [p_i, H] = [p_i, q_j] \frac{\partial H}{\partial q_j} = - \delta_{ij} \frac{\partial H}{\partial q_j}
= \frac{\partial H}{\partial q_i} \, .
												\end{align}
We now combine the canonical coordinates $q_i$ and the momenta $p_i$ into a new set of $N$ generalized coordinates:
\be
\label{13}
                                 z     = (q_1, \ldots, q_{N/2}, p_1, \ldots, p_{N/2}) \, .
		  								       \ee
	Then the canonical $P.B.$ from (\ref{9}) can be very elegantly written with the aid of the Poisson tensor $\omega$ as
	\be
	\label{14}
			[z_a, z_b] = \omega_{ab} \, , \quad \quad    det (\omega_{ab}) \neq 0 \quad    (1 \, \mbox{for canonical coordinates})  	            							
\ee
with
\begin{align}
\label{15}
\omega_{ab} = \,\,\,
  \begin{blockarray}{cccccc}
           & \scriptstyle{q_1}       & \scriptstyle{p_1}       & \scriptstyle{q_2}    & \scriptstyle{p_2}    & \cdots \\
    \begin{block}{c(ccccc)}
    \scriptstyle{q_1}    & [q_1,q_1] & [q_1,p_1] & \cdots & \cdots & \cdots \\
    \scriptstyle{p_1}    & [p_1,q_1] & [p_1,p_1] & \cdots & \cdots & \cdots \\
    \scriptstyle{q_2}    &           &           & \ddots &        &        \\
    \scriptstyle{p_2}    &           &           &        & \ddots &        \\
    \vdots &           &           &        &        & \ddots \\
    \end{block} 
  \end{blockarray} 
\,\,\,=\,\, 
\begin{pmatrix}
 \dboxed{\begin{matrix} 0 & 1 \\ -1 & 0 \end{matrix}} &  & \\
 & \dboxed{\begin{matrix}0 & 1 \\ -1 & 0 \end{matrix}} & \\
 & & \ddots
\end{pmatrix}
\end{align}

or
\be
\label{16}
\omega_{ab} =   \mathrm{diag} \left[ \lk \begin{array}{cc} 
                                 0 & 1 \\
                                 - 1 & 0
                                \end{array} \right) \, , \ldots
                                \lk
                                \begin{array}{cc} 
                                 0 & 1 \\
                                 - 1 & 0
                                \end{array} \right) \right] \, .
								\ee
The inverse of $\omega_{ab} , \omega^{ab} = -\omega_{ba}$, is defined according to
\be
\omega_{ac} \omega^{cb} = \delta_a^b \nonumber
\ee
\be
\mbox{e.g., for} \quad N = 2, \quad  (q_1, p_1):
\lk  \begin{array}{cc}
      0 & 1\\
      - 1 & 0
     \end{array} \right) 
     \lk  \begin{array}{cc}
      0 & - 1\\
      1 & 0
     \end{array} \right)
     \lk  \begin{array}{cc}
      1 & 0\\
      0 & 1
     \end{array} \right)
\, . \nonumber
\ee
Then our $P.B.$ (\ref{8}) can now be written in abbreviated form as
\be
\label{17}
	[A (z), B (z)]_{P.B.} = \frac{\partial A}{\partial z_a} [z_a, z_b] \frac{\partial B}{\partial z_b} = \omega_{ab}
	\frac{\partial A}{\partial z_a} \frac{\partial B}{\partial z_b} = \partial^a A \omega_{ab} 
	\partial^b B \, .
	\ee
	Accordingly, the Hamiltonian equations are given by
\begin{align}
 \label{18}
 \dot{z}_a &= [z_a, H] = \omega_{ab}  \partial^b H \, , \quad \quad \omega_{ab} \, \mbox{is independent of} \, z \nonumber\\
 \lk {\dot{q}  \atop \dot{p}} \right) &= \lk \begin{array}{cc}
                                             0 & 1 \\
                                             - 1 & 0 
                                            \end{array} \right) 
                                            \lk \begin{array}{c}
                                                 \frac{\partial H}{\partial q} \\
                                                 \frac{\partial H}{\partial p}
                                                \end{array} \right) \quad : \quad \dot{q} = 
                                                \frac{\partial H}{\partial p} \, , \quad \dot{p} = - \frac{\partial H}{\partial q} \, .
\end{align}
We call a transformation from the original variables $(q_i, p_i)$ to new variables $(Q_i, P_i)$
\begin{align}
 \label{19}
 Q_i &= Q_i (q_i, p_i) \, , \nonumber\\
 P_i &= P_i (q_i, p_i) \, ,
\end{align}
\underline{canonical}, if the fundamental $P.B.$ for the new $(Q_i, P_i)$ also are of form (\ref{9}):
\be
\label{20}
[Q_i, Q_j] = 0 = [P_i, P_j] \, , \quad \quad [Q_i, P_j] = \delta_{ij} \, .
\ee
It can be easily shown that as a consequence of formulas (\ref{20}), 
the Jacobi determinant of a canonical transformation is equal to one (''the canonicity is preserved''): 
\be
\label{21}
	\mathrm{det}(\omega_{qp}) = \mathrm{det} (\omega_{QP}) = 1 \, .
	\ee
To prove (\ref{21}), we use $N = 2$. With the help of (\ref{10}), the following simply applies:
\be
\label{22}
[Q, P]_{q, p} = \lk \frac{\partial Q}{\partial q} \frac{\partial P}{\partial p} - 
\frac{\partial Q}{\partial p} \frac{\partial P}{\partial q} \right) \, .
\ee
If $Q$ and $P$ are to be a pair of canonical conjugate variables, then according to equations (\ref{20}), $[Q,P] = 1$ is valid; thus
\be
[Q, P] = \left| \begin{array}{cc}
                 \frac{\partial Q}{\partial q} & \frac{\partial Q}{\partial p} \\ 
\frac{\partial P}{\partial q} &  \frac{\partial P}{\partial p} 
                \end{array}\right|  = 1\, . \nonumber
\ee
The main characteristic of a canonical transformation is, however, that in the new variables, the canonical equations also hold:
the Hamiltonian canonical equations are form invariant (covariant) under (\ref{19}), 
but with a new Hamiltonian function $K$:
\be
\label{23}
H (q (Q, P), p (Q, P)) = K    (Q, P) \, .
\ee
We show this again for one degree of freedom:
\begin{align}
 \dot{Q} = [Q, H]_{q, p} &= \frac{\partial Q}{\partial q} \frac{\partial H}{\partial p} - 
 \frac{\partial Q}{\partial p} \frac{\partial H}{\partial q} \nonumber\\
 &= \frac{\partial Q}{\partial q} \lk \frac{\partial H}{\partial Q} \frac{\partial Q}{\partial p} + 
 \frac{\partial H}{\partial P} \frac{\partial P}{\partial p} \right) - 
 \frac{\partial Q}{\partial p} \lk \frac{\partial H}{\partial Q}  \frac{\partial Q}{\partial q}
 + \frac{\partial H}{\partial P}  \frac{\partial P}{\partial q} \right) \nonumber\\
 & = \frac{\partial H}{\partial P} \lk \frac{\partial Q}{\partial q} \frac{\partial P}{\partial p} - 
 \frac{\partial Q}{\partial p} \frac{\partial P}{\partial q} \right) = 
  \frac{\partial H}{\partial P} [Q, P] \equiv   \frac{\partial K}{\partial P} (Q, P) \, .
\nonumber
\end{align}
Similarly for $\dot{P}$.
\bi

\no
The literature on Hamiltonian dynamics is dominated by canonical transformations. But this an unfortunate situation, 
since it is often impossible to introduce convenient  variables which are also canonical. In a ``pseudocanonical''
transformation, the new variables must not be canonical, i.e., they need \underline{not} satisfy (\ref{20}). 
However, the transformation equations, as before, have to be invertible.
\bi

\no
For non-canonical phase-space variables, 
instead of (\ref{18}), we now write
\be
\dot{z}_a = [z_a, H] = \sigma_{ab} (z) \partial^b H \, , \quad \quad a, b = 1, 2, \ldots N \,  \nonumber
\ee
with $\sigma_{ab} (z)$ instead of (\ref{14})
\be
\label{24}
[z_a, z_b] = \sigma_{ab} (z) \, .
\ee 
Properties of $\sigma_{ab} (z)$
\be
\label{25}
[z_a, z_b] = - [z_b, z_a] : \quad \quad \sigma_{ab} (z) = - \sigma_{ba} (z) \, .
\ee
Inverse
\be
\label{26}
\sigma_{ac} (z) \sigma^{cb} (z) = \delta^b_a \, .
\ee
From the Jacobi identity
\be
[z_a, [z_b, z_c]] + [z_b, [z_c, z_a]] + [z_c, [z_a, z_b]] = 0 \nonumber
\ee
follows as a condition for the $\sigma$'s:
\be
\label{27}
\partial_a \sigma_{bc} (z)  + \partial_b \sigma_{ca} (z) + \partial_c \sigma_{ab} (z) = 0 \, .
\ee
If we change the phase-space coordinates according to the transformation
   \be
   z \to         z'(z) \nonumber
   \ee
we get as $L.B.$ for the new coordinates
\begin{align}
 \sigma'_{ab} &= [z'_a, z'_b] = [z'_a (z), z'_b (z)] \nonumber\\
 &=\frac{\partial z'_a}{\partial z_c} [z_c, z_d] \frac{\partial z'_b}{\partial z_d} = 
 \frac{\partial z'_a}{\partial z_c} \sigma_{cd} (z) \frac{\partial z'_b}{\partial z_d} \, . \nonumber
\end{align}
Therefore						
\be
\label{28}
\sigma'_{ab} = \frac{\partial z'_a}{\partial z_c} \frac{\partial z'_b}{\partial z_d} \sigma_{cd} \, .
\ee
Here the $\sigma_{ab}$ turn out to be components of a two-rank contravariant tensor.
\bi

\no
Moreover, we already know that the $L.B.$ for two arbitrary phase-space functions $A(z)$, $B(z)$ can be written as
\begin{align}
 \label{29}
 [A (z), B (z)] &= \frac{\partial A}{\partial z_a} [z_a, z_b] \frac{\partial B}{\partial z_b} \, , \nonumber\\
 &= \frac{\partial A}{\partial z_a} \sigma_{ab} \frac{\partial B}{\partial z_b} \, .
\end{align}
If $z_a =  (q_1, \ldots, q_{N/2}, p_1, \ldots, p_{N/2})$ are canonical coordinates, then we get 
for the transformation $z \to z'$   ($z'$ is not necessarily canonical)
\be
\sigma'_{ab} = \frac{\partial {z'}_a}{\partial z_c} \frac{\partial {z'}_b}{\partial z_d} \omega_{cd} \, . \nonumber
\ee
Here, $\omega_{ab}$ is again the ``flat metric'' in the canonical $q-p$ basis (\ref{16}).
\bi

\no													
We now want to illustrate the practicality of the noncanonical coordinates using three very important examples from 
physics. Let us begin with the mathematical pendulum. The energy (not Hamiltonian) of a particle in the Earth's 
gravitational field is given in right-angle Cartesian coordinates by
\be
\label{30}
H' (x, p_x, p_y) = \frac{1}{2m} (p^2_x + p^2_y) - mgx .
\includegraphics[width=.2\textwidth]{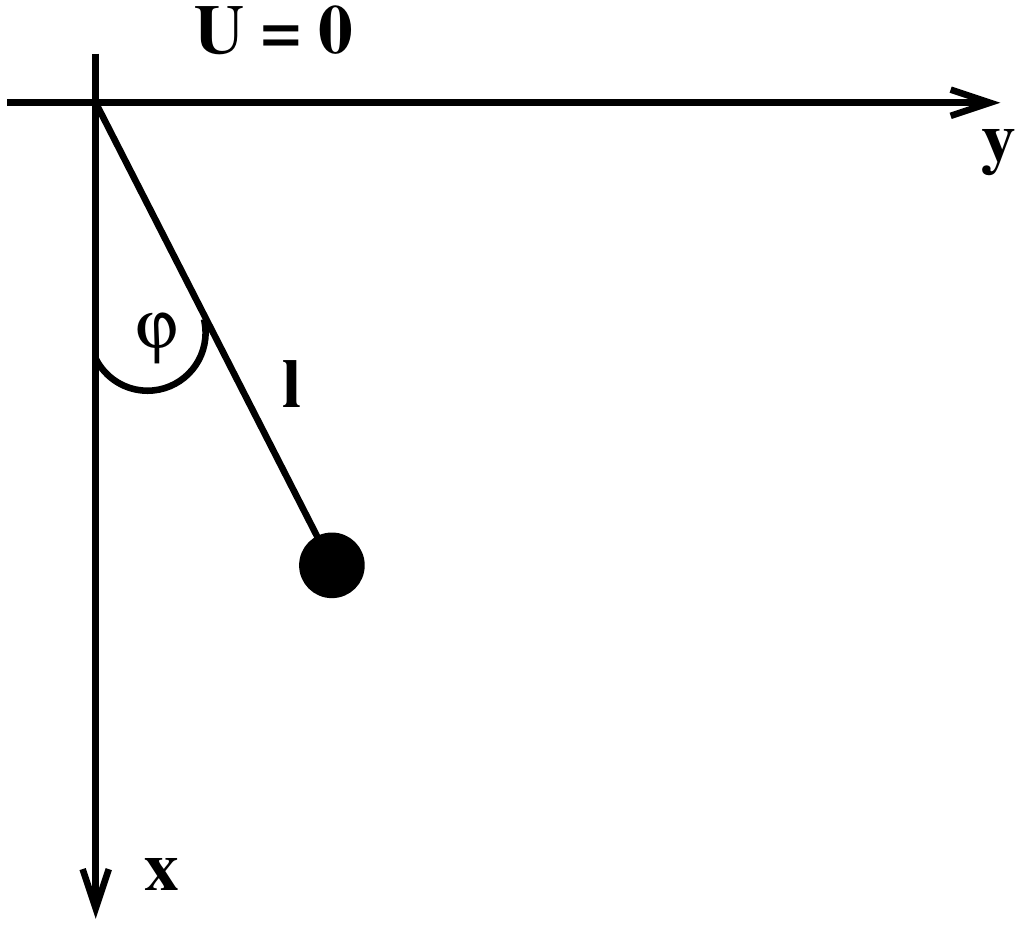}
\ee
but subject to the constraint 
\be
\label{31}
	x^2 + y^2 = l^2 \, .            					           \ee
It should be observed that the variables in (\ref{30}) 
are 
noncanonical, so that they do not satisfy the conventional Hamiltonian canonical equations of motion, because of the additional
constraint (\ref{31}). It is, however, well known that we can express $H'$ in canonical coordinates
$(\varphi, p_\varphi)$ with the resultant Hamiltonian:
			\be
			\label{32}
			H (\varphi, p_\varphi) = \frac{1}{2 m l^2} p^2_\varphi - m g l \cos \varphi \, .
\ee
Hence we can express the equations of motion with a single degree of freedom. The associated phase space 
$(\varphi, p_\varphi)$ is two dimensional and endowed with a Poisson bracket structure:
\be
\label{33}
[\varphi, \varphi]_{P. B} = 0 = [p_\varphi, p_\varphi] \, , \quad \quad [\varphi, p_\varphi]_{P. B} = 1 \, .
	\ee
This can also be formulated by saying that the symplectic structure of this two-dimensional phase space is given by the symplectic matrix
\be
\label{34}
\omega = 
\bordermatrix{
  & \varphi	& p_\varphi   \cr
\varphi & 0 & 1 \cr
p_\varphi & - 1 & 0 \cr
} \, .
\ee
The two-dimensional matrix representation is typical for every canonical pair $(q,p)$.
Our next goal is to enlarge the so far two-dimensional phase pace into four dimensions with pseudo-canonical variables 
$(r, p_r, \varphi, p_\varphi)$. We call them pseudo-canonical variables because, as will be shown, $r$
and $p_r$ satisfy certain constraints and hence cannot be canonical variables. This means that there are no Poisson brackets for the 
$(r, p_r)$ variables. Instead, the Lie brackets take over and are the building elements of the symplectic matrix.
Here are the details. We begin with the pseudo-Hamiltonian    
\be
\label{35}
	H' (r, p_r, \varphi, p_\varphi)	= \frac{1}{2 m} \lk p^2_r + \frac{1}{r^2} p^2_\varphi \right)	- m g r \cos \varphi \, .
	\ee
The pseudo-canonical transformation is given by
\be
(x, p_x, y, p_y ) \to (r, p_r, \varphi, p_\varphi) \nonumber
\ee
with
\begin{align}
 x &= r \cos \varphi \, , \quad \quad p_x = p_r \cos \varphi - \frac{p_\varphi}{r} \sin \varphi \, , \nonumber\\
 y &= r \sin \varphi \, , \quad \quad p_y = p_r \sin \varphi + \frac{p_\varphi}{r} \cos \varphi \, . \nonumber
\end{align}
Inverting these equation yields
\begin{align}
\cos \varphi &= \frac{x}{r} \, , \quad \quad p_r = p_x \cos \varphi + p_\varphi \sin \varphi \, , \nonumber\\
 \sin \varphi &= \frac{y}{r} \, , \quad \quad p_\varphi = - p_x r \sin \varphi + p_y  r \cos \varphi \, . \nonumber
\end{align}
Here are our two constants of motion (constraints):
\be
\label{36}
(x^2 + y^2) = r^2 \, , \quad \quad \frac{x p_y + y p_y}{(x^2 + y^2)^{1/2}} = p_r \, .
\ee
These quantities are called Casimirs and are constants of motion because of the form of the bracket rather 
than the form of the Hamiltonian. In $(\varphi, p_\varphi)$
space, of course, the brackets (as well as the bracket matrix $\omega$) are canonical.
Before introducing the constraints (\ref{36}), the four-dimensional matrix $\omega'$ is given by
\begin{align}
 \label{37}
 \omega' = \begin{array}{c}
 \\
            r \\
            p_r \\
            \varphi \\
            p_\varphi 
           \end{array}          
                      \lk
           \begin{array}{cccc}
            r & p_r & \varphi & p_\varphi 
                       \\
            \left[r, r\right] & \left[r, p_r\right] & \left[r, \varphi \right] & \left[r, p_\varphi \right]                    
            \\
            \left[p_r, r\right] & \left[p_r, p_r\right] & \ldots \ldots &  \ldots \ldots \\
            \vdots && \vdots & \\
            \left[p_\varphi, r\right] & \ldots & \left[p_\varphi, \varphi\right] & \left[p_\varphi, p_\varphi\right] 
           \end{array}
           \right) 
                                 \begin{array}{l}
            \mbox{constraints} \\
            \longrightarrow \\
            \left[r, r \right] = 0 \\
            \mbox{etc.}
           \end{array}
           \lk \begin{array}{cc}
                O_2 & O_2 \\
                O_2 &  \lk \begin{array}{cc}
                       0 & 1 \\
                       - 1 & 0
                      \end{array}
\right) 
               \end{array}
\right) \, .
\end{align}
After taking into account the two constraints in (\ref{36}), 
this four-dimensional $\omega'$ structure is reduced to a two-dimensional phase subspace $(N = 2)$
with canonical coordinates $(\varphi, p_\varphi)$ and the well-known symplectic matrix $\lk \begin{array}{cc}
                       0 & 1 \\
                       - 1 & 0
                      \end{array}
\right)$, meaning one degree of freedom, so that our dynamical 
system ``mathematical pendulum'' is soluble.
\bi

\no
When we use the one-dimensional harmonic oscillator in the $x$ representation, we have for the Schr\"odinger Hamiltonian
\be
H (x, \partial_x) =  - \frac{\hbar^2}{2 m} \partial^2_x + \frac{1}{2} m \omega^2 x^2 \quad \mbox{with} \quad
[\partial_x, x] = 1 \, , \quad [\partial^2_x, x] =  2 \partial_x \, . \nonumber
\ee
To set up the corresponding Lie algebra we define the following ``operators'':
\be
L_+ = \frac{1}{2} x^2 \, , \quad \quad L_- = - \frac{1}{2} \partial^2_x \, , \qquad \quad L_3 = \frac{1}{2} x \partial_x + \frac{1}{4} 
\nonumber
\ee
so that the pseudo-Hamiltonian can be written as
\be
H' = \frac{\hbar^2}{m} L_- + m \omega^2 L_+ \, . \nonumber
\ee
Now it is easy to   prove that
\be
\left[L_+, L_- \right] = 2 L_3 \, , \quad \quad \left[L_3, L_\pm\right] = L_\pm \, . \nonumber
\ee
This is the $SO(3)$ Lie algebra of angular momenta, which is locally isomorphic to the $SU(2)$ Lie algebra
and thus our problem is naturally soluble. 
\bi

\no
Our next example is the description of a charged particle in an external constant magnetic field, i.e., with the Hamiltonian function
\be
\label{38}
H (\vq, \vp) = \frac{1}{2 m} \lk \vp - \frac{e}{c} \vA (\vq) \right)^2 \, , \quad \quad	
\vA (\vq) = \frac{1}{2} \vB \times \vq \, .
\ee
This choice of vector potential (gauge) guarantees that $\vB$ indeed is constant, because
\begin{align}
 \vB &= \vec{\nabla}_{\vq} \times \vA (\vq) = \frac{1}{2} \vec{\nabla} \times (\vB \times \vq) \nonumber\\
 &= \frac{1}{2} 
 \Big[ (\vq \cdot \vec{\nabla}) \vB - 
  (\vB \cdot \vec{\nabla}) \vq + \vB (\vec{\nabla} \cdot \vq) - 
 \vq (\vec{\nabla} \cdot B) \Big]  
 \nonumber\\
 &= \frac{1}{2} \left[ 0 - \vB + 3 \vB - 0 \right] = \frac{1}{2} (- \vB + 3 \vB) = \vB \, . \nonumber
\end{align}
Herewith we get as Hamiltonian equations of motion
\begin{align}
 \label{39}
 \dot{p}_i &= - \frac{\partial H}{\partial q_i} = \frac{e}{mc} \lk p_j  - \frac{e}{c} A_j (\vq) \right) 
 \frac{\partial}{\partial q_i} A_j (\vq) \, , \quad \quad A_j (\vq) = \frac{1}{2} \epsilon_{jkl} B_k q_l \nonumber\\
 &= \frac{e}{mc} \lk p_j - \frac{e}{c} A_j (\vq) \right) \frac{1}{2} \epsilon_{jki} B_k \nonumber\\
 &= \frac{e}{2 mc} \left[ \lk \vp - \frac{e}{c} \vA (\vq) \right) \times \vB \right]_i \\
 \label{40}
 \dot{q}_i &= \frac{\partial H}{\partial p_i} = \lk \vq - \frac{e}{c} \vA (\vq) \right)_i \, .
 \end{align}
Now we consider the noncanonical transformation
\begin{align}
(\vq, \vp)   & \longrightarrow (\vx', \vec{v}') \equiv \vz'_a \nonumber
\\
               & \mbox{noncanonical coordinates} \nonumber
\end{align}
so:  
\begin{align}
 \label{41}
 \vx' &= \vq \nonumber\\
 \vv' &= \frac{1}{m} \lk \vp - \frac{e}{c} \vA (\vq) \right) \equiv
 \frac{1}{m} \vec{\Pi} \, \mbox{(noncanonical, gauge invariant)} \, .
\end{align}
$
(\vx', \vv')$ parametrize the phase-space just as well as $(\vq, \vp)$.
Now it holds for the fundamental Lie brackets:
\begin{align}
 \label{42}
 {\vv'}_i &= q_i : \quad \quad \left[{\vx'}_i, {x'}_j \right]_{q, p} =   
 \frac{\partial {x'}_i}{\partial q_k} \frac{\partial x'_j}{\underbrace{\partial p_k}_{= 0}}  -
 \frac{\partial x'_i}{\underbrace{\partial p_k}_{= 0}} 
 \frac{\partial {x'}_j}{\partial q_k} = 0
 \nonumber\\
 {v'}_i &= \frac{1}{m} \lk p_i - \frac{e}{c} A_i (q_j) \right): \quad \quad \left[{x'}_i, {v'}_j \right]_{q, p} = 
 \frac{1}{m} \delta_{ij} = 
 \left[{v'}_i, {x'}_j \right]_{q, p} \nonumber\\
 A_i (\vq) &= \frac{1}{2} \epsilon_{ijk} B_j q_k
 \nonumber\\
 \left[{v'}_i, {v'}_j \right]_{q, p} &= \frac{e}{m^2 c} \lk \frac{\partial A_j}{\partial q_i} - \frac{\partial A_i}{\partial q_j} \right) = 
 \frac{e}{m^2c} B_{ij} = \frac{1}{m^2} \left[ \Pi_i, \Pi_j \right] \nonumber\\
 & \hspace{5cm} \frac{e}{c} B_{ij} = F_{ij} \nonumber\\
 &= \frac{e}{m^2 c} \epsilon_{ijk} B_k = \frac{1}{m} \Omega_{ij} \nonumber\\
 \overset{\leftrightarrow}{\Omega} &= \Omega_{ij} = \frac{e}{m c} \epsilon_{ijk} B_k \, .
\end{align}
So we have found:
\begin{align}
 \label{43}
 \sigma'_{a b} &= \frac{1}{m} \lk \begin{array}{cc}
                                   O_3 & 1_3\\
                                   - 1_3 &  \overset{\leftrightarrow}{\Omega}
                                  \end{array}
                                  \right) \\
                                  \label{44}
                                  \sigma'^{ab} &= m 
\lk \begin{array}{cc}
                                   \overset{\leftrightarrow}{\Omega} & - 1_3\\
                                   1_3 & O
                                  \end{array}
                                  \right) \, ,
\end{align}	
so that indeed
\begin{align}
 \sigma'_{ac} \sigma'^{cb} &= \lk \begin{array}{cc}
                                   0 & 1\\
                                   -1 &  \overset{\leftrightarrow}{\Omega}
                                  \end{array}
                                  \right) 
                                  \lk \begin{array}{cc}
                                    \overset{\leftrightarrow}{\Omega} & - 1\\
                                   1 & 0
                                  \end{array}
                                  \right) \nonumber\\
                                  &= \lk \begin{array}{cc}
                                   1 & 0\\
                                   0 & 1
                                  \end{array}
                                  \right) = 1 \, . \nonumber
\end{align}
In canonical coordinates $(\vq, \vp)$:									
\be
\label{45}
H (\vq, \vp) = \frac{1}{2 m} \lk \vp - \frac{e}{c} \vA (\vq) \right)^2 \, .
\ee
In non-canonical coordinates 
\be
\label{46}
(\vx', \vv'): H' (\vx', \vv') = \frac{m}{2}  \vv'^2  = \frac{1}{2 m} \vec{\Pi}^2				\, .
\ee
Non-canonical equations of motion:
\be
\frac{d z'_a}{d t} = [z'_a, H'] = \sigma'_{ab} \frac{\partial H'}{\partial z'_b}\nonumber
\ee
or
\begin{align}
\label{47}
 \frac{d}{dt} \lk {\vx' \atop \vv'} \right) &= \frac{1}{m} 
\lk \begin{array}{cc} 
     0 & 1 \\
     - 1 &  \overset{\leftrightarrow}{\Omega}
         \end{array}\right)
    \lk {\frac{\partial H'}{\partial\vx'} \atop \frac{\partial H'}{\partial \vv'}} \right) = 
    \lk
    \begin{array}{c}
     \vv' \\
     \frac{e}{m c} \epsilon_{ijk} B_k v'_j
    \end{array}
    \right) \nonumber\\
    &= \lk
    \begin{array}{c}
     \vv' \\
     \frac{e}{mc} \vv' \times \vB
    \end{array}
    \right) 
\end{align}
\be
\label{48}
\mbox{Newton-Lorentz equation.} \atop (\mbox{No} \, \vA \, , \mbox{gauge invariant!}) \, .
\ee
One should note that in $H'(\vx', \vv') = \frac{m}{2} \vv'^2 = \frac{1}{2m} \vec{\Pi}^2$, 
the nonphysical magnetic vector potential $\vA$ has disappeared ! ? 
The components of the $\sigma'_{ab}$
tensor can be used to calculate the $L.B.$ of two phase-space functions $f (\vx', \vv'), g (\vx',\vv')$:
\be
[f, g] = \frac{\partial f}{\partial {z'}_a} \sigma'_{ab} \frac{\partial g}{\partial {z'}_b} \, . \nonumber
\ee
The result follows directly from (\ref{43})
\begin{align}
 \label{49}
 [f, g] &= \frac{\partial f}{\partial \vx'} 
 \mathclap{\raisebox{3.5ex}{\scalebox{3}[1]{$\frown$}}}
\mathclap{\raisebox{-3.5ex}{\scalebox{3}[1]{$\smile$}}}
 \frac{\partial f}{\partial \vv'} \frac{1}{m} 
 \lk \begin{array}{cc}
      0 & 1\\
      - 1 & \overset{\leftrightarrow}{\Omega}
     \end{array}
     \right)
      \lk \begin{array}{c}        
      \frac{\partial g}{\partial \vx'}\\
      \frac{\partial g}{\partial \vv'}
     \end{array}
     \right) \nonumber\\
     &= \frac{1}{m} 
\frac{\partial f}{\partial \vx'} 
 \mathclap{\raisebox{3.5ex}{\scalebox{3}[1]{$\frown$}}}
\mathclap{\raisebox{-3.5ex}{\scalebox{3}[1]{$\smile$}}}
\frac{\partial f}{\partial \vv'}  
\begin{pmatrix}
 \frac{\partial g}{\partial \vv'} \\
 - \frac{\partial g}{\partial \vx'} 
 +  \overset{\leftrightarrow}{\Omega} \cdot \frac{\partial g}{\partial \vv'}
\end{pmatrix} \nonumber\\
&= \frac{1}{m} \lk
\frac{\partial f}{\partial \vx'} \cdot \frac{\partial g}{\partial \vv'} - \frac{\partial f}{\partial \vv'} \cdot 
\frac{\partial g}{\partial \vx'} \right) + \frac{e}{m^2 c} \vB \cdot 
\lk \frac{\partial f}{\partial \vv'} \times \frac{\partial g}{\partial \vv'} \right) 
\, ,
\end{align}
which for
\be
f = \Pi_i (\vv') = m v'_i \, , \quad \quad g = \Pi_j (\vv') = m v'_j \, \mbox{reproduces} \, (\ref{42}). \nonumber
\ee
The $\vB$ or, respectively, the $\vA$ field, which disappeared in $H' = \frac{m}{2} \vv'^2$, 
can be rediscovered in the altered symplectic structure $\sigma'$ (\ref{43}) of the gauge invariant  pseudo-Hamiltonian system
$H' = \frac{1}{2 m} \vec{\Pi}^2$.
\bi

\no
A further important example of the application of noncanonical coordinates can be found in force-free rigid bodies. 
Here $\dot{\vL}  = 0$ holds in the inertial system. Relative to the fixed-body reference system, the \underline{same}
angular-momentum vector has the components $\{K_i\}_{i = 1, 2, 3} = \vL^{body}$.
\bi

\no
Now we perform a non-canonical transformation
\begin{align}
\begin{array}{lll}
(\phi, \theta, \psi, p_\phi, p_\theta, p_\psi)  &  \longrightarrow  & (K_1, K_2, K_3) \nonumber\\
\mbox{6-dimensional canonical phase space} & &   \mbox{3-dimensional reduced noncanonical phase space} 
\end{array} \nonumber
\end{align}
The
formation equations are
\begin{align}
 K_1 &= \lk p_\phi \frac{1}{\sin \theta}  - p_\psi \cot \theta \right) \sin \psi + p_\theta \cos \psi \, , \nonumber\\
  K_2 &= \lk p_\phi \frac{1}{\sin \theta}  - p_\psi \cot \theta \right) \cos \psi - p_\theta \sin \psi \, , \nonumber\\
  K_3 &= p_\psi \, . \nonumber
\end{align}
With the help of fundamental canonical brackets $[\phi, p_\phi], \ldots$ one can prove the following validity for the fundamental $L.B.$:
\begin{align}
\label{50}
	[K_1, K_2] &  = -K_3 \nonumber\\
	[K_2, K_3] &  = -K_1 \quad \quad \mbox{Note the minus signs!} \nonumber\\
	[K_3, K_1] & = -K_2 
\end{align}
                  or     
                  \be
                  \label{51}
                  [K_\alpha, K_\beta] = - \epsilon_{\alpha \beta \gamma} K^\gamma \, ,
                  \ee
                                 i.e.,
                                 \be
                                 \sigma^{\alpha \beta} = - \epsilon^{\alpha \beta \gamma} K_\gamma \nonumber
\ee
                  or				                                                                         									  
\be
\label{52}
\sigma = \lk \begin{array}{ccc}
              0 & - K_3 & K_2 \\
              K_3 & 0 & - K_1 \\
              - K_2 & K_1 & 0
             \end{array} \right) \, .
\ee
Herewith one can form the (noncanonical) $L.B.$ of two arbitrary functions $A (\vK), B (\vK)$ according to
\begin{align}
 \label{53}
 [A (\vK), B (\vK)] &= \frac{\partial A}{\partial K^\alpha} [K^\alpha, K^\beta] \frac{\partial B}{\partial K^\beta} \nonumber\\
 &= - \epsilon^{\alpha \beta \gamma} K_\gamma \frac{\partial A}{\partial K^\alpha} \frac{\partial B}{\partial K^\beta} \nonumber\\
 &= - \vK \cdot \lk \frac{\partial A}{\partial \vK} \times \frac{\partial B}{\partial \vK} \right) \, .
\end{align}
With the pseudo-Hamiltonian function with $I_i$ the principal axis of inertia
\be
H' = \frac{K^2_1}{2 I_3} + \frac{K^2_2}{2 I_2} + \frac{K^2_3}{2 I_3}
\nonumber
\ee
follows for the equations of motion:									
\be
\label{54}
\frac{d}{dt} K^\alpha = [K^\alpha, H'] = \sigma^{\alpha \beta} \frac{\partial H'}{\partial K^\beta} \, ,
\ee
i.e.,
\be
\frac{d}{dt} \lk \begin{array}{c}
K^1\\ K^2 \\ K^3
                                 \end{array}
\right) = \lk
\begin{array}{ccc}
 0 & - K_3 & K_2 \\
 K_3 & 0 & - K_1 \\
 - K_2 & K_1 & 0
\end{array} \right) 
\lk 
\begin{array}{c}
 \frac{\partial H'}{\partial K_1} \\
 \frac{\partial H'}{\partial K_2} \\
 \frac{\partial H'}{\partial K_3} 
\end{array} \right) \, , \quad \quad
\frac{\partial H'}{\partial K^\beta} = \frac{K_\beta}{I_\beta} \, . \nonumber 
\ee
For example:
\be
\label{55}
\frac{d}{dt} K_1 = K_3 \frac{K_2}{I_2} + K_2 \frac{K_3}{I_3} \, ,
\ee
  i.e.
  \begin{align}
  \dot{K}_1 = & \lk \frac{1}{I_3}  - \frac{1}{I_2} \right) K_2 K_3 \nonumber\\
\mbox{and cyclic order} &. \nonumber               
   \end{align}
These are Euler's equations of motion for the force-free rigid body.
With $K_i = I_i \omega_i (i = 1,2,3)$ we obtain the well-known formula for the force-free rigid body
\be
\label{56}
(I_i - I_j) \omega_i \omega_j - \sli^3_{k = 1} \epsilon_{ijk} I_k \dot{\omega}_k = 0 \, .
\ee
It is remarkable that without further calculation our system proves to be soluble. This follows from the Casimir
\be
C= K^2_1 + K^2_2 +  K^2_3 \, , \nonumber
\ee
which is conserved by the form of the brackets. 
Thus the motion of a free rigid body actually lives in a two-dimensional state space on the sphere $C = constant$, 
equivalent to a one-degree-of-freedom system, and therefore the motion is integrable.
\bi

\no
To show how useful our former results are, let us calculate the non-relativistic propagation function for a charged particle 
in three dimensions in presence of a constant $B$ field in the $z$ direction. 
\bi

\no
We rewrite our findings (\ref{47}) in terms of the pseudo-momentum $\vec{\Pi} = m \vv$:
\begin{align}
 \label{57}
 \frac{d \vx (t)}{dt} &= \frac{\vec{\Pi}}{m} (t) \, , \nonumber\\
 \frac{d \vec{\Pi} (t)}{dt} &= \frac{e}{c}  \overset{\leftrightarrow}{\Omega} \cdot \vec{\Pi} (t) \, , \quad \quad
 ( \overset{\leftrightarrow}{\Omega})_{ij} = 
 \frac{e}{mc} \epsilon_{ijk} B^k \, .
\end{align}
From now on we choose the axis of the magnetic field in the $z$ direction so that our pseudo-Hamiltonian $\cH$ reads:
\be
\label{58}
\cH = \frac{1}{2m} (\Pi^2_1 + \Pi^2_2) + \frac{P^2_3}{2m} = \cH_\perp + \frac{P^2_3}{2m} \, .
\ee
The free-particle propagation in the third direction can be easily treated, so we drop it and introduce
\be
\vr (t) = \lk {x_1 (t) \atop x_2 (t)} \right) \, , \quad \quad \vec{\Pi} (t) = 
\lk {\Pi_1 (t) \atop \Pi_2 (t)} \right) \, . \nonumber
\ee
We leave it as an exercise for the reader to show that in the now two-dimensional problem the matrix $\sigma$ is given by
\begin{align}
\label{59}
\sigma_{ab} = \frac{1}{m} 
\lk \begin{array}{cc}
     0_2 & 1_2 \\
     - 1_2 & \frac{e B}{mc} \overset{\leftrightarrow}{\epsilon}
    \end{array}
    \right) \, , \quad \quad 
    (\overset{\leftrightarrow}{\epsilon})_{ab} = 
  \lk \begin{array}{cc}
     0 & 1 \\
     - 1 & 0
    \end{array}
    \right) \, , \quad \quad  
    ( \overset{\leftrightarrow}{\epsilon})^{ab} = 
  \lk \begin{array}{cc}
     0 & -1 \\
      1 & 0
    \end{array}
    \right)  & = - \epsilon_{ab} \, , \nonumber\\
    & \epsilon_{ac} \epsilon^{cb} = \delta_a^b \, .
\end{align}
Hence the non-canonical equations of motion take the form
\begin{align}
 \label{60}
 \lk \dot{\bar{x}} \atop \dot{\vv} \right) &= \frac{1}{m} 
 \lk \begin{array}{cc}
     0 & 1 \\
      -1 & \frac{e B}{m c} \overset{\leftrightarrow}{\epsilon}
    \end{array}
    \right) 
 \lk \begin{array}{c}
     0  \\
      \frac{\partial \cH_\perp}{\partial \vv}
    \end{array}
    \right)    = \frac{1}{m}
    \lk \begin{array}{cc}
     0_2 & 1_2 \\
      -1 & \frac{e B}{m c} 
       \overset{\leftrightarrow}{\epsilon}
          \end{array}
    \right) 
    \lk \begin{array}{c}
     0  \\
      m \vv
    \end{array}
    \right)     
    \\
    \label{61}
    & \mbox{or} \quad \frac{d \vr}{d t} = \frac{\vec{\Pi}}{m}  \\
    \label{62}
     \frac{d \vec{\Pi}}{d t} & = \frac{e B}{mc} 
      \overset{\leftrightarrow}{\epsilon}
     \cdot \vec{\Pi} \, , \quad \quad 
    \Omega = \frac{e B}{m c} \quad \mbox{cyclotron frequency} \nonumber\\
    & = \Omega 
         \overset{\leftrightarrow}{\epsilon}
\cdot \vec{\Pi} \, .
\end{align}
The solution of (\ref{62}) for $\vec{\Pi}$ is needed for determining the explicit expression for $\cH_\perp (t) = \frac{\vec{\Pi}^2}{2m}$.                
This information enables us to calculate the three-dimensional non-relativistic Feynman propagation function. 
Solving equations (\ref{61}) and (\ref{62}) for $\vr (t)$ and $\vec{\Pi} (t)$, is equivalent to solving the following Lie
bracket relations
\begin{align}
 \label{63}
 \left[ r_i (t), \Pi^2 (t) \right] &= 2 \Pi_i (t) \, , \nonumber\\
 \left[ \Pi_i (t), \Pi^2 (t) \right] &= 2 \frac{e}{c} B_{ij} \Pi^j (t) \nonumber\\
 &= F_{ij} \Pi^j = 2 \frac{e}{c} \epsilon_{ij3} B^3 \Pi^j (t) \, .
\end{align}
On the way to calculating the Feynman propagation function we could use Schwinger's proper time method. But instead of doing so, we prefer
another method, making use of the classical action and the Van Vleck determinant.
\bi

\no
To begin with, we remember from the harmonic oscillator the classical action \cite{1}
\be
S_{el} = \frac{m \omega}{2 \sin (\omega \tau)} \left[ (x^2 + {x'}^2) \cos (\omega \tau) - 2 x x' \right] \, . \nonumber
\ee
Using the Van Vleck determinant
\be
D = \mathrm{det} \left[ - \lk \frac{\partial^2 S_{el}}{\partial x \partial x'} \right) \right] = \frac{m \omega}{\sin (\omega \tau)} : \quad \quad
\sqrt{D} = \sqrt{\frac{m \omega}{\sin (\omega \tau)}} \nonumber
\ee
we obtain for the propagation function of the linear harmonic oscillator
\begin{align}
 K (x, x'; \tau) &= \sqrt{\frac{1}{2 \pi i \hbar}} \sqrt{D} e^{\frac{i}{\hbar} S_{el}} \nonumber\\
 &= \lk \frac{m \omega}{2 \pi i \hbar \sin (\omega \tau)} \right)^{1/2} \exp 
 \left\{ \frac{i m \omega}{2 \hbar \sin (\omega \tau)} \left[ (x^2 + x'^2) \cos (\omega \tau) - 2 x x' \right] \right\} \, . \nonumber
\end{align}
If we now follow the same path for a charged particle gauge fixed for a magnetic field in $z$ direction we obtain 
                                                                                                                  \begin{align}
 \label{64}
 \lk \vr \right. & = \left. (x_1, x_2) \, , \quad {\vr}' = ({x'}_1, {x'}_2) \, , \quad  (\overset{\leftrightarrow}{\epsilon})_{ij} =                                                                                                         
 \lk \begin{array}{cc}
 0 & 1 \\
 - 1 & 0
           \end{array} \right) \right) \nonumber\\
           S_{el} &= \frac{m}{2} \left\{ \frac{\omega}{2} \cot (\omega \tau) (\vr - \vr')^2 + \omega \vr \cdot 
           \overset{\leftrightarrow}{\epsilon} \cdot 
           \vr' \right\} \, , \quad \omega = \frac{e B}{m c} \, .
\end{align}
The results of the two space derivatives we need for the Van Vleck determinant are given by
\begin{align}
 & - \frac{m}{2} \omega \cot \lk \frac{\omega \tau}{2} \right) 1 + \frac{m}{2} \omega \overset{\leftrightarrow}{\epsilon} = - \frac{m}{2}
 \frac{\omega}{\sin \lk \frac{\omega \tau}{2} \right)} \lk \cos \lk \frac{\omega \tau}{2} \right) 1 - 
\overset{\leftrightarrow}{\epsilon}\sin \lk \frac{\omega \tau}{2} \right) \right) \nonumber\\
 &= - \frac{m}{2} \frac{\omega}{\sin \lk \frac{\omega \tau}{2} \right)} \lk \cos \lk \frac{\omega \tau}{2} \right) 1  - 
 i \sigma_2 \sin \lk \frac{\omega \tau}{2} \right) \right) = - \frac{m}{2} \frac{\omega}{\sin \lk \frac{\omega \tau}{2} \right)}
 e^{- i \sigma_2 \lk \frac{\omega \tau}{2} \right)} \, . \nonumber
\end{align}
From here we obtain the determinant
\be
D = \lk - \frac{m}{2} \frac{\omega}{\sin \lk \frac{\omega \tau}{2} \right)} \right)^2 : \quad \quad \sqrt{D} = 
\frac{m}{2} \frac{\omega}{\sin \lk \frac{\omega \tau}{2} \right)} \, . \nonumber
\ee
The full result for the propagation function in a gauge that fixes the magnetic field in the third direction is then given by 
($\omega = \frac{eB}{m c}$, cyclotron frequency)
\begin{align}
 \label{65}
 K (\vr, \vr'; \tau) K (x_3, x'_3; \tau) &= \frac{1}{2 \pi  i \hbar} \frac{m}{2} \frac{\omega}{\sin \lk \frac{\omega \tau}{2} \right)}
 e^{\frac{i}{\hbar} \frac{m}{2} \left[ \frac{\omega}{2} \cot \lk \frac{\omega \tau}{2} \right) 
 (\vr - \vr')^2 + \omega \vr \cdot \vec{\epsilon} \cdot \vr' \right]} \nonumber\\
 & \cdot \sqrt{\frac{m}{2 \pi i \hbar \tau}} e^{\frac{i}{\hbar} \frac{m}{2} \frac{(x_3 - x'_3)^2}{\tau}} \, .
\end{align}
It might be interesting to write in $K (\vr, \vr'; \tau)$: 
($L$ indicates a straight path connecting $\vr'$ with $\vr$):
 \begin{align}
 \label{66}
 e^{\frac{i}{\hbar} e \il^{\vr}_{\vr'} d \vec{\zeta} \cdot \vA (\vec{\zeta})} 
 &= 
 e^{\frac{i}{\hbar} \frac{m}{2} \omega {\vr} \cdot \vec{\epsilon} \cdot {\vr'}} 
 \nonumber\\   
 &= e^{\frac{i}{\hbar} 
 \frac{m}{2} \frac{e B}{m c} (x_1 {x'}_2 - x_2 {x'}_1)} 
 \nonumber\\
 &= e^{\frac{i}{\hbar}  \frac{e}{2 c} \phi} \, , \quad \quad \phi  = A r e a \cdot B \, .
\end{align}
Hence we can also write
\be
\label{67}
K (\vr, \vr'; \tau) = \frac{1}{2 \pi i \hbar} \frac{m}{2} \frac{\omega}{\sin \lk \frac{\omega \tau}{2} \right)} 
e^{\frac{1}{\hbar} e \il^{\vr}_{\vr'_L} d \vec{\zeta} \cdot \vA (\vec{\zeta})}
e^{\frac{i}{\hbar} \frac{m}{2} \frac{\omega}{2} \cot \lk \frac{\omega \tau}{2} \right) (\vr - \vr')^2} \, , 
\ee
an expression we will meet again in the full relativistic case of a charged particle travelling in a 
constant electromagnetic field. 
\bi

\no
So let us turn to the calculation of the propagation function using the relativistic coordinate representation
of $K (x', s; x'', 0)$. One can think of $s$ as   of proper time introduced by V. Fock and J. Schwinger \cite{2}. 
We will not adopt this method but instead follow the former non-relativistic strategy, i.e., making use of the 
Van Vleck determinant in $d = 4$ dimensions. We also set $\hbar = 1$.
\bi

\no
As before, we begin with a Lagrangian, in our case with a matrix-valued Lagrangian, set up
the equations of motion, and compute the classical action. So let us start with \cite{3}
\be
\label{68}
L   = \frac{1}{4} \dot{x}_\mu \dot{x}^\mu + e A^\mu \dot{x}_\mu + \frac{e}{2} \sigma^{\mu \nu} F_{\mu \nu} (x (s)) \, .
\ee
The associated (``pseudo-'') Hamiltonian is clearly $\lk \Pi = \frac{\dot{x}}{2} = (p - eA) \right)$
\begin{align}
 H &= p \dot{x} - L = \lk \frac{\dot{x}}{2} + e A \right) \dot{x} - \frac{1}{4} \dot{x}^2 - e A \dot{x} - \frac{e}{2} \sigma F = 
 \frac{\dot{x}^2}{4} - \frac{e}{2} \sigma F \nonumber\\
 \mbox{or} \quad H &= \Pi^2 - \frac{e}{2} \sigma F \, . \nonumber
\end{align}
From (\ref{68}) we obtain the equations of motion:
\begin{align}
 \frac{d}{ds} \lk \frac{\partial L}{\partial \dot{x}^\mu} \right) - \frac{\partial L}{\partial {x}^\mu} &= 0 : \quad \quad
 \frac{\partial L}{\partial \dot{x}^\mu} = \frac{1}{2} \dot{x}_\mu + e A_\mu \, , \nonumber\\
 & \hspace{2cm} \frac{\partial L}{\partial x^\mu} =  e \frac{\partial A_\lambda}{\partial x^\mu} \dot{x}^\lambda + \frac{e}{2}
 \sigma^{\lambda \nu} \frac{\partial F_{\lambda \nu}}{\partial x^\mu} \nonumber\\
  \frac{d}{ds} \lk \frac{1}{2} \dot{x}_\mu + e A_\mu \right) &= e \frac{\partial A_\lambda}{\partial x^\mu} \dot{x}^\lambda + \frac{e}{2} 
  \sigma^{\lambda \nu} \frac{\partial F_{\lambda \nu}}{\partial x^\mu} \, , \nonumber\\
  \frac{1}{2} \ddot{x}_\mu + e \frac{\partial A_\mu}{\partial x^\lambda} \dot{x}^\lambda : \quad \quad \frac{1}{2} \ddot{x}_\mu &= e \lk 
  \frac{\partial A_\lambda}{\partial x^\mu } - \frac{\partial A_\mu}{\partial x^\lambda} \right) \dot{x}^\lambda + 
  \frac{e}{2} \sigma^{\lambda \nu} \frac{\partial F_{\lambda \nu}}{\partial x^\mu} \, . \nonumber
\end{align}
This provides us with the equation of motion
\be
\label{69}
\ddot{x}_\mu = 2 e F_{\mu \nu} \dot{x}^\nu + e \sigma^{\lambda \nu} \frac{\partial F_{\lambda \nu}}{\partial x^\mu} \, .
\ee
Using
\be
\dot{x}_\mu = 2 \Pi_\mu \nonumber
\ee
we have
\be
\dot{\Pi}_\mu = \frac{1}{2} \ddot{x}_\mu = e F_{\mu \nu} \dot{x}^\nu + \frac{e}{2} \sigma^{\lambda \nu} \frac{\partial F_{\lambda \nu}}{\partial x^\mu}
\nonumber
\ee
or
\be
\label{70}
\dot{\Pi}_\mu = 2 e F_{\mu \nu} \Pi^\nu + \frac{e}{2} \sigma^{\lambda \nu} \frac{\partial F_{\lambda \nu}}{\partial x^\mu} \, .
\ee
From now on we consider constant fields only, i.e., $F_{\mu \nu} = const.$. Since we want to calculate the classical action, i.e.,
\be
\label{71}
S_{el} = \il^s_0 d \lambda L (\dot{x} (\lambda), x (\lambda))
\ee
we first have to integrate (in matrix notation)
\be
\ddot{x} = 2 e F \dot{x} \, , \nonumber
\ee
which can be solved with the Ansatz
\be
x (\lambda) = e^{2 e F \lambda} \dot{x}  (0) \nonumber
\ee
and a further integration yields
\be
x (\lambda) - x (0) = \il^\lambda_0 d \lambda' e^{2 e F \lambda'} \dot{x} (0) = \frac{1}{2 e F} \left[ e^{2 e F \lambda} - 1 \right] 
\dot{x} (0) \nonumber
\ee
or with the initial condition $x (\lambda = 0) = x''$ and $x (\lambda = s) = x'$ we obtain
\be
x' - x'' = \frac{1}{2 e F} \left[ e^{2 e F s} - 1 \right] \dot{x} (0) \nonumber
\ee
or
\be
\dot{x} (0) = \frac{1}{e^{2 e F s} - 1} 2 e F (x' - x'') \, . \nonumber
\ee
So far we have
\begin{align}
 \label{72}
 x (\lambda) - x (0) &= \frac{1}{2 e F} \left[ e^{2  e F \lambda} - 1 \right] \frac{1}{e^{2 e F s} - 1} 2 e F (x' - x'') \nonumber\\
 &= \frac{e^{2 e F \lambda} - 1}{e^{2 e Fs} - 1} (x' - x'') \, .
\end{align}
We need
\be
\label{73}
S = \il^s_0 d \lambda \left\{ \frac{1}{4} \dot{x}^T (\lambda) \dot{x} (\lambda) + e \dot{x}^T (\lambda) A (\lambda) \right\}
+ \frac{e}{2} s \sigma F \, .
\ee
In the present case ${g (F)^T} = g (F^T) = g (- F)$. 
Therefore with
\begin{align}
 \dot{x} (\lambda) & {=} 
 e^{2 e F \lambda} \dot{x} (0)  {=} \frac{e^{2 e F \lambda}}{e^{2 e Fs} - 1} 
 2  e F (x' - x'')
 \nonumber\\
 \dot{x}^T (\lambda) \dot{x} (\lambda) &= (x' - x'')^T \frac{(- 2 e F)}{e^{- 2 e Fs} - 1} \cancel{e^{- 2 e F \lambda}} 
 \cancel{e^{2 e F \lambda}}  \frac{2  e F}{e^{2 e F s} - 1} 2 e F (x' - x'') \nonumber\\
&=  (x' - x'')^T \frac{(- 2 e F)}{\cancel{e^{- e Fs}} \lk e^{- e Fs} - e^{e Fs} \right)} 
\frac{(2 e F)}{\cancel{e^{e Fs}} \lk e^{e Fs} - e^{- eFS} \right)} (x' - x'') \nonumber\\
&= (x' - x'')^T \frac{e F}{\sin h (e Fs)} \frac{e F}{\sin h (e Fs)} (x' - x'') \, , \nonumber\\
\to \frac{1}{4} \dot{x}^T (\lambda) \dot{x} (\lambda) &= \frac{1}{4} (x' - x'') ^T \frac{(e F)^2}{\sin h^2 (e Fs)} (x' - x'') \nonumber
\end{align}
we obtain for the first term in (\ref{73})
\be
\label{74}
\frac{1}{4} \il^s_0 d \lambda \dot{x} (\lambda)^T \dot{x} (\lambda)  = \frac{1}{4} (x' - x'')^T \frac{(eF)^2 s}{\sin h^2 (e Fs)} (x' - x'') \, .
\ee
Now we turn to the expression 
\begin{align}
 \label{75}
 \il^s_0 d \lambda \{ e \dot{x}^T A \} &= \il^{x'}_{x''} d x^T (e A) \nonumber\\
 &= e  \il^{x'}_{x''}  d x^T (e A + \frac{1}{2} F \cdot (x - x'')) - 
 e \frac{1}{2} \il^{x'}_{x''} d x^T F \cdot (x - x'') \, .
\end{align}
The first integral on the right-hand-side of (\ref{75}) is independent of the path of 
integration, since the curl of the integrand 
vanishes: 
\begin{align}
 & \partial_\mu \lk A_\nu + \frac{1}{2} F_{\nu \lambda} (x - x'')^\lambda \right) - 
 \partial_\nu \lk A_\mu + \frac{1}{2} F_{\mu \lambda} (x - x'')^\lambda \right) \nonumber\\
 &= \partial_\mu A_\nu - \partial_\nu A_\mu + \frac{1}{2} F_{\nu \mu} - \frac{1}{2} F_{\mu \nu} \nonumber\\
 &= F_{\mu \nu} - F_{\mu \nu} = 0 \, . \nonumber
\end{align}
Hence instead of integrating along the classical path $x (\lambda)$ we can integrate along the straight path
connecting $x'$ and $x''$:
\be
\label{76}
q (\lambda) = x'' + (x' - x'') \frac{\lambda}{s} \, , \quad \quad q (0) = x'' \, , \quad \quad
q (s) = x'
\ee
\begin{align}
 & e \il^{x'}_{x''} d x^T \lk A + \frac{1}{2} F \cdot (x - x'') \right) \nonumber\\
 &= e \il^{x'}_{x''} d  q^T A + \frac{e}{2} \il^{x'}_{x''} d  q^T F \cdot (q - x'') \nonumber\\
 &= e \il^{x'}_{x''} d  q^\mu A_\mu (q) + \frac{e}{2} \il^s_0 \frac{d \lambda}{s} 
 \underbrace{(x' - x'') F (x' - x'')}_{(x' - x'')^\mu F_{\mu \nu} \cdot (x' - x'')^\nu = 0 \atop F_{\mu \nu} = 
 - F_{\nu \mu}} \frac{\lambda}{s} \nonumber\\
 &= e \il^{x'}_{x''} d q^\mu A_\mu (q) \, . \nonumber
\end{align}
Finally we need the second term in (\ref{75})
\be
\label{77}
- \frac{e}{2} \il^{x'}_{x''} d x^T F \cdot (x - x'') = \frac{e}{2} \il^{x'}_{x''} d \lambda 
\cdot{x}^T F \cdot (x - x'') \, .
\ee
\begin{align}
 \dot{x}^T F (x - x'') &= (x - x'')^T \frac{(- 2 e F)}{e^{- 2 e Fs} - 1} 
 e^{- 2 e F \lambda} F \cdot (x - x'') \nonumber\\
 &= (x - x'')^T \frac{(- 2 e F)}{e^{- 2 e Fs} - 1} e^{- 2 e F \lambda} F 
 \frac{e^{2 e F \lambda} - 1}{e^{2 e F s} - 1} (x' - x'') \nonumber\\
 &= (x' - x'')^T \frac{(- 2 e F)}{\cancel{e^{- e Fs}} (e^{- e Fs} - e^{e Fs})} F 
 \frac{1 - e^{- 2 e F \lambda}}{\cancel{e^{e Fs}} (e^{e Fs} - e^{- Fs})} (x' - x'') \nonumber\\
 ./. &= (x' - x'')^T \frac{e F^2}{\sin h e Fs} \cdot \frac{1}{2} 
 \frac{(1 - e^{- 2 e F \lambda})}{\sin h e Fs} (x' - x'') \nonumber\\
 \il^s_0 d \lambda ./. &= \frac{1}{2} (x' - x'')^T \frac{e F^2}{\sin h^2 e Fs} 
 \left[ s + \frac{1}{2 e F} \lk e^{- 2e Fs} - 1 \right) \right] (x' - x'') \nonumber\\
 &= \frac{1}{2} (x' - x'')^T \left[ \frac{e F^2 s}{\sin h^2 e Fs} + \frac{e F^2}{\sin h^2 e Fs} 
 \frac{1}{2 e F} e^{- e Fs} \lk e^{- e Fs} - e^{e Fs} \right) \right] (x' - x'') \nonumber\\
 &=  \frac{1}{2} (x' - x'')^T  \left[ \frac{e F^2 s}{\sin h^2 e Fs}  - \frac{e F^2}{\sin h^2 e F s} \frac{e^{- e FS}}{e F} \sin h e Fs 
 \right] (x' - x'') \nonumber\\
 &= \frac{1}{2} (x' - x'')^T \left[ \frac{e F^2 s}{\sin h^2 e Fs} - e^{- e Fs} \frac{F}{\sin h e Fs} \right] (x' - x'') \nonumber
\end{align}
So the contribution of the integral is given by
\be
\label{78}
- \frac{e}{2} \il^{x'}_{x''} d x^T F \cdot (x - x'') = - \frac{1}{4} (x' - x'')^T \left[ \frac{(e F)^2 s}{\sin h^2 e Fs} - 
\frac{e^{- e Fs} (e F)}{\sin h e Fs} \right] (x' - x'') \, .
\ee
The first term in (\ref{78}) cancels the result (\ref{74}). So we obtain for the classical action 
\begin{align}
 S (x', x''; s) & \underset{\text{($\ref{13}$)}}{=}  e \il^{x'}_{x''} d q^\mu A_\mu (q) + \frac{1}{4} (x' - x'')^T
 e^{- e Fs} \frac{eF}{\sin h e Fs} (x' - x'') + \frac{e}{2} s \sigma F \nonumber\\
 &= e \il^{x'}_{x''} d q_\mu  A^\mu (q) + \frac{1}{4} (x' - x'')^T e F 
 \underbrace{\frac{\cos h e Fs - \sin h e Fs}{\sin h e Fs}}_{(\cot h e F s - \cancel{1})_{(x' - x'') F (x' - x'') = 0}} 
 (x' - x'') + \frac{e}{2} \sigma F s \nonumber\\
 &= e \il^{x'}_{x''} d q_\mu A^\mu (q) + \frac{1}{4} (x' - x'')^T e F \cot h (e Fs) (x' - x'') + \frac{e}{2} \sigma_{\mu \nu}
 F^{\mu \nu} s \, .\nonumber
\end{align}
Then the final result reads
\be
\label{79}
S (x', x''; s) = e \il^{x'}_{x''} d q_\mu A^\mu (q) + \frac{1}{4} (x' - x'')^\alpha e F_\alpha^\beta [\cot h (e Fs)]_\beta^\gamma (x' - x'')_\gamma
+ \frac{e}{2} \sigma_{\mu \nu} F^{\mu \nu} s \, .
\ee
With this information we are able to compute the propagation function 
by a classical WKB approximation which is exact for the constant-field case
\begin{align}
 \label{80}
 K (x', s; x'', 0) &= \il_{x (0) = x'' \atop x (s) = x'} [d x (\lambda)] e^{i S [x (\lambda)]} \nonumber\\
 S &= \il^s_0 d \lambda L (x, \dot{x}) = (\ref{79})
\end{align}
or
\be
\label{81}
K (x', s; x'', 0) = (2 \pi i)^{- \frac{n}{2}} \sqrt{D} e^{i S}
\ee
with the Van Vleck determinant
\be
\label{82}
D (x', x''; s)^{n = 4} = (- 1)^4 \det \lk \frac{\partial^2 S}{\partial x'^\mu \partial x''^\nu} \right) \, .
\ee
We begin with
\begin{align}
 \label{83}
 & \frac{\partial^2}{\partial {x'}^\mu \partial {x''}^\nu} \left[ \frac{1}{4} (x' - x'')^\alpha  e F_\alpha^\beta 
 ( \cot h e F s)_\beta^\gamma 
 (x' - x'')_\gamma \right] \nonumber\\
 &= \frac{\partial}{\partial x''^\nu} \left[ \frac{1}{4} e F_\mu^\beta (\cot h e Fs)_\beta^\gamma (x' - x'')_\gamma 
 + \frac{1}{4} (x' - x'')^\alpha e F_\alpha^\beta (\cot h e Fs)_{\beta \mu} \right] \nonumber\\
 &= \frac{\partial}{\partial x''^\nu} \left[ \frac{1}{2} (x' - x'')^\alpha e F_\alpha^\beta (\cot h e Fs)_{\beta \mu} \right] \nonumber\\
 &= - \frac{1}{2} e F_\nu^\beta (\cot h e Fs)_{\beta \mu} \, .
 \end{align}
Furthermore we have to compute
\begin{align}
\label{84}
\frac{\partial}{\partial {x'}^\mu} \il^{x'}_{x''} d q^\alpha A_\alpha (q) &= \frac{\partial}{\partial {x'}^\mu} \il^s_0 d \lambda
\frac{\partial q^\alpha}{\partial \lambda} A_\alpha (q) \nonumber\\
&= \il^s_0 d \lambda \left[ \lk \frac{\partial}{\partial \lambda} \frac{\partial q^\alpha}{\partial {x'}^\mu} \right) A_\alpha (q) + 
\frac{\partial q^\alpha}{\partial \lambda} \frac{\partial A_\alpha (q)}{\partial q^\beta} \frac{\partial q^\beta}{\partial {x'}^\mu} \right]
\nonumber\\
&= \left. \frac{\partial q^\alpha}{\partial {x'}^\mu} A_\alpha (q) \right]^{\lambda = s}_{\lambda = 0} + \il^s_0 d \lambda
\left[ - \frac{\partial q^\alpha}{\partial {x'}^\mu} \frac{\partial A_\alpha}{\partial q^\beta} \frac{\partial q^\beta}{\partial \lambda}
+ \frac{\partial q^\alpha}{\partial \lambda} \frac{\partial A_\alpha (q)}{\partial q^\beta} 
\frac{\partial q^\beta}{\partial {x'}^\mu} \right]\nonumber\\
q (s) = x' \atop q (0) = x'' &= A_\mu (x') + \il^s_0 d \lambda \frac{\partial q^\alpha}{\partial \lambda} 
\left[ \frac{\partial A_\alpha}{\partial q^\beta} - \frac{\partial A_\beta}{\partial q^\alpha} \right] 
\frac{\partial q^\beta}{\partial {x'}^\mu} 
\nonumber\\
q (x', x'', \lambda) &= x'' + (x' - x'') \frac{\lambda}{s} : \quad \frac{\partial q^\alpha}{\partial \lambda} = \frac{(x' - x'')^\alpha}{s}
\nonumber\\
& \hspace{5cm} \frac{\partial q^\beta}{\partial {x'}^\mu} = \delta_\mu^\beta \frac{\lambda}{s} \nonumber\\
\frac{\partial}{\partial {x'}^\mu} \il^{x'}_{x''} d q^\alpha A_\alpha (q) &= A_\mu (x') + \il^s_0 d \lambda
(x' - x'')^\alpha \frac{1}{s} F_{\beta \alpha} \delta_\mu^\beta \frac{\lambda}{s} \nonumber\\
&= A_\mu (x') + \frac{1}{2} F_{\mu \alpha} (x' - x'')^\alpha \, .
 \end{align}
   Therefore
 \be
 \label{85}
 \frac{\partial^2}{\partial x''^\nu \partial x'^\mu} \il^{x'}_{x''} d q^\alpha A_\alpha (q) = - \frac{1}{2} F_{\mu \nu} \, .
 \ee
 Together with (\ref{83}) we obtain
 \begin{align}
  \label{86}
  \frac{\partial' S}{\partial x'^\mu \partial x''^\nu} &= - \frac{1}{2} e F_{\mu \nu} - \frac{1}{2} e
  F_\nu^\beta (\cot h e Fs)_{\beta \mu} \nonumber\\
  &= - \frac{1}{2s} \left[ e Fs + e Fs (\cot h e Fs) \right]_{\mu \nu} \nonumber\\
  &= - \frac{1}{2s} \left[ e Fs \frac{\sin h e Fs + \cos h e Fs}{\sin h e Fs} \right]_{\mu \nu} \nonumber\\
   &= - \frac{1}{2s} \left[ e Fs \frac{e^{e Fs}}{\sin h e Fs} \right]_{\mu \nu} \, .
 \end{align}
Given (\ref{86}), the Van Vleck determinant is given by
\begin{align}
 \label{87}
 n = 4: \quad D  &= \det \left\{ (-1) \frac{1}{2s} e Fs \frac{e^{e Fs}}{\sin h e Fs} \right\} \nonumber\\
 &= - 1 \frac{1}{16 s^4} \det \lk e F s \frac{e{Fs}}{\sin h e F s} \right) \nonumber\\
 &= -  1 \frac{1}{16 s^4} e^{tr \ln \lk \frac{e Fs}{\sin h e Fs} e^{e Fs} \right)} \nonumber\\
 &= - \frac{1}{16 s^4} e^{- tr \ln \frac{\sin h e Fs}{e Fs} + es \overbrace{tr F}^{= 0}} \nonumber\\
 \sqrt{D} &= \frac{i}{4 s^2} e^{- \frac{1}{2} tr \ln \frac{\sin h e Fs}{e Fs}} \, .
\end{align}
We finally end up with
\begin{align}
  K (x', s; x'', 0)  &= - \frac{1}{4 \pi^2} \frac{i}{4 s^2} e^{- \frac{1}{2} tr \ln \frac{\sin h e Fs}{e Fs}} e^{i S} \nonumber
\end{align}
or
\begin{align}
 \label{88}
   K (x', s; x'', 0)  &= - \frac{i}{16 \pi^2} \frac{1}{s^2} e^{- \frac{1}{2} tr \ln \lk \frac{\sin h e Fs}{e Fs} \right)} 
  \nonumber\\
  & \cdot e^{i e \il^{x'}_{x''} d q_\mu A^\mu (q)} e^{i \left[ \frac{1}{4} (x' - x'') \lk e F \cot h e Fs \right) 
  (x' - x'') \right]} e^{i \frac{e}{2} \sigma F s} \, \\
  & \mbox{holonomy factor} \quad \quad \mbox J.S., P.R. {\bf 82}, 664 (51) \nonumber\\
  & \mbox{counter gauge factor} \quad \quad (3.20) \nonumber\, .
\end{align}
This result should be compared with Julian Schwinger's superb article, Physical Review {\bf 82}, 664 (1951) [ref. 2], formula (3.20).
A useful result is then given by (c.f. \ref{80})
\begin{align}
 \label{89}
 & \il_{x (0) = x'' \atop x (s) = x'} \left[d x (\lambda) \right] e^{i \il^s_0 d \lambda \left[ \frac{1}{4} 
 \dot{x}^2 (\lambda) + e A (x (\lambda)) \dot{x} (\lambda) \right]} \nonumber\\
 &= \underbrace{e^{i e \il^{x'}_{x''} d q_\mu A^\mu (q)}}_{\mbox{gauge dependence} \atop \mbox{isolated!}} 
 \frac{- i}{(4 \pi)^2} \frac{1}{s^2} e^{- \frac{1}{2} tr \ln
 \lk \frac{\sin h e Fs}{e Fs} \right)} \nonumber\\
 & \cdot e^{i \frac{1}{4} (x' - x'') (e F \cot h e Fs) (x' - x'')} \, .
 \end{align}
It takes a little more to show that (\ref{88}) makes its appearance when calculating the Green's function of an 
electron in an external constant electromagnetic field (in an arbitrary function gauge):
\begin{align}
 \label{90}
 G (x, x' | A ) &= \phi (x, x' |  A) \frac{1}{(4 \pi)^2} \il^\infty_0 \frac{ds}{s^2} 
 \left[ m - \frac{1}{2} \gamma^\mu [f (s) + e F]_{\mu \nu} (x - x')^\nu \right] \nonumber\\
& \cdot e^{- i m^2 s - L (s)} \exp \left[ \frac{1}{4} (x - x') f (s) (x - x') \right]e^{i \frac{e}{2} \sigma Fs} \, . 
\end{align}
Here we have used the abbreviations
\be
f (s) = e F \cot h (e Fs) \quad \mbox{and} \quad L (s) = \frac{1}{2} tr \ln \frac{\sin h (e Fs)}{e Fs} \, , \nonumber
\ee
and
\be
\phi (x, x' | A) = \exp \left\{ i e \il^x_{x'} d \zeta_\mu \left[ A^\mu (\zeta) + \frac{1}{2} F^{\mu \nu} (\zeta - x')_\nu \right]
\right\} \nonumber
\ee
carries completely the gauge dependence of the propagation function. Needless to say, with (\ref{90}) in our hands, we can compute several 
important processes in QED, as there are the effective Lagrangian, in particular the one-loop effective Lagrangian, the pair-production rate, the Heisenberg-Euler
Lagrangian, photon-photon scattering, the axial vector anomaly, dispersion effects for low-frequency photons and many more \cite{3}.
\bi

\no
Also note that in our last relativistic problem we never used operator quantum field theory.
We always dealt with measured physical quantities like, for instance, the charge and mass of the electron,
and nowhere were we confronted with any kind of renormalisation procedure.

\end{document}